\documentclass[12pt]{iopart}
\usepackage{graphicx}
\usepackage{hyperref}
\usepackage{amsmath}
\usepackage{comment} 

\def\4He{$^4$He}
\newcommand{\rhos}{\ensuremath{\rho_s}}

\newcommand{\bv}[1]{\ensuremath{\mathbf{#1}}}
\newcommand{\dd}{\ensuremath{\operatorname{d}}}

\begin{document}
\title[Turbulent coupling in superfluid acoustics.]{Turbulent dissipative coupling in nanoscale multimode superfluid acoustics.}
\author{F Novotný$^1$, M Talíř$^1$, and E Varga$^1$}
\address{$^1$ Faculty of Mathematics and Physics, Charles University, Ke Karlovu 3, 121 16 Prague,
Czech Republic}
\ead{filip.novotny@mff.cuni.cz}
\ead{emil.varga@matfyz.cuni.cz}

\begin{abstract}
Superfluid helium, the inviscid low-temperature phase of liquid \4He, enables investigation of flows with reduced dimensionality since, due to the vanishing viscosity, sub-micron flow channels can be constructed. In such strongly confined volumes filled with superfluid, the longitudinal acoustic wave is a coupled fluctuation of pressure and entropy density called fourth sound. In this work, we use multiple 4th sound acoustic modes inside a nano-superfluidic acoustic resonator in a pump-probe arrangement to observe localized clusters of quantized vortices leading to two-dimensional turbulence. The localised turbulence enables controllable and asymmetric dissipative coupling between acoustic modes. Furthermore, we derive a general procedure for analytically estimating the superfluid acoustic resonance frequencies inside a volume with mechanically compliant walls. Our work confirms earlier assumptions that turbulence in similar nanofluidic systems initially develops in localized areas of high shear. The multimode pump-probe methods presented here will allow future experiments to study the dynamics of two-dimensional quantum turbulence, e.g., the free decay.
\end{abstract}
\maketitle


\section{Introduction}
Superfluid \4He is attracting renewed interest as an acoustic medium due to its low acoustic dissipation and the ability to fill nearly arbitrary volumes. This allows significant design freedom for prototype small-footprint detectors of gravitational waves \cite{vadakkumbatt2021} or dark matter \cite{hirschel2024,hertel2019}; or as the mechanical element in optomechanical systems \cite{shkarin2019,spence2023,childress2017,brown2023,spence2021}.

Apart from the ordinary (or first) sound, superfluid helium supports several unique sound-like modes due to the two-fluid nature of superfluid flows above approximately 1~K, which can be decomposed into inviscid superflow, which transports superfluid density and no entropy, and viscous normal flow, which transports both normal fluid density and entropy \cite{tilley2003}. The second sound is a wave of temperature or entropy density, and the fourth sound is a coupled wave of pressure and entropy density arising in strongly confined geometries where normal fluid can become viscously clamped, which leads to the oscillation of the superfluid component alone\cite{tilley2003}. The vorticity (i.e., the curl of the velocity field) in the superfluid component is restricted to the cores of discrete quantized vortices \cite{tilley2003}, each carrying a single quantum of circulation. These vortices can form a complex, chaotic tangle (quantum turbulence \cite{barenghi2023}). In a quasi-2D system, vortices will span the height of the fluid layer and will act essentially as vortex points moving on a plane, which is a system known as Onsager vortex gas \cite{onsager1949,sachkou2019,simula2014,Gauthier2019}.

The nonclassical superfluid sounds are attenuated in the presence of quantized vortices, which can be used to study stationary turbulence in quasi-2D flows of high aspect ratio \cite{varga2020, novotny2024, novotny2024_arxiv}. Flows where the horizontal length scales are several orders of magnitude larger than the vertical (which are geometrically comparable to, e.g., large-scale atmospheric or oceanic flows \cite{hogstrom1999, rosell2015}) display many features of true two-dimensional turbulence, notably the inverse cascade \cite{boffetta2012,alexakis2023}. In classical liquids, due to the finite viscosity, experiments are constrained to thin soap films or magnetohydrodynamic fluids typically with free surface \cite{couder1984, vorobieff1999, fang2017}. In superfluid helium, thanks to the vanishing viscosity, high flow aspect ratios can be achieved in fully enclosed channels. Flows with typical heights of several 100~nm and aspect ratios over 1000 were used to probe forced quasi-2D turbulence \cite{varga2020, novotny2024, novotny2024_arxiv}, where complex multistable and hysteretic behaviour was observed at the transition to turbulence. However, the spatial distribution of the quantized vortices had to be assumed.

In this work, we demonstrate a first step toward a spatially-resolved acoustic probe of two-dimensional turbulence in nanoscale flow using multimode superfluid acoustics in a fourth sound resonator. In particular, we confirm the assumption from past works \cite{varga2020, novotny2024_arxiv} that the turbulence is localised in areas of maximal flow velocity near sharp edges. This also lends support to the localised nucleation of vortices observed in $^3$He-A in similar geometries \cite{shook2024}. We show that localised turbulence can be generated by one acoustic mode and probed by another via increased dissipation. This turbulence-mediated dissipative coupling also opens new possibilities for the study of the dynamics of two-dimensional turbulence and could allow, e.g., in-situ tuning of mechanical dissipation in superfluid optomechanical systems.

The paper is organized as follows: In section \ref{sec:setup}, we describe the acoustic resonator and experimental methods; in section \ref{sec:modes}, we show the detected acoustic resonances and their assignment to simulated modes, and in section \ref{sec:modes-calculation} we derive analytically the resonant frequencies of 4th sound acoustic modes in a compliant container; in section \ref{sec:coupling} we then characterize the interaction between resonant modes mediated by localized turbulent clusters of quantized vortices, after which conclusions follow.

\section{Experimental setup}
\label{sec:setup}
\begin{figure}[h!]
    \centering
    \includegraphics[width=\textwidth]{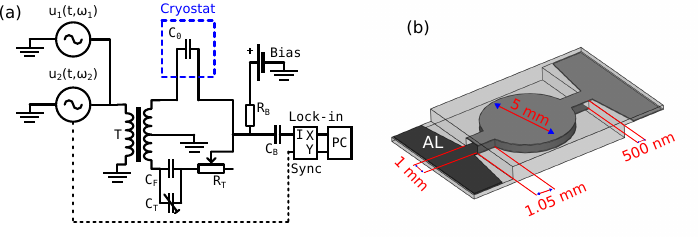}
    \caption{(a) Measurement circuit used for the detection of the 4$^{th}$ sound resonance. $u_1$ and $u_2$ are the driving voltages, while the lock-in is referenced to one of them. (b) Sketch of the chip used in the experiment with dimension. The height of the nanofluidic cavity is not to scale.}
    \label{fig:circuit}
\end{figure}

The resonant motion of the superfluid component of He~II was studied using a Helmholtz resonator, a well-established device for the realization of a 2D-confined He~II flow \cite{varga2020, novotny2024, novotny2024_arxiv, rojas2015, souris2017},. Helmholtz resonator is a fused silica chip with a fully-enclosed nanofluidic cavity. The cavity consists of a circular basin 5~mm in diameter, which is connected to the surroundings, bulk He~II, by two channels with width 1~mm and length 1.05~mm, see Fig.\ref{fig:circuit}(b). Two 50~nm thick aluminium electrodes are evaporated on the surfaces of the basin, which serve to drive and capacitively sense the resonant motion of the fluid. The Helmholtz resonator used in this work had cavity height $h \approx$ 500~nm and geometry similar to previous works \cite{varga2020, novotny2024, novotny2024_arxiv}.

To drive and detect the resonant motion of the superfluid inside the cavity, the device was connected to one arm of the capacitance bridge (see Fig.\ref{fig:circuit}(a)) similarly as in \cite{varga2021}, which was tuned before the data acquisition so that approximately zero current was detected by the lock-in amplifier. The bridge was excited by a pair of function generators (voltages $u_1$ and $u_2$), one of which was used for the lock-in reference (which corresponded to the detection peak, see below). Applied AC-voltage $u(t)$ at the resonance frequency $\omega$ causes a deformation of the chip capacitor plates, which results in a change of its capacitance $C_0 + \delta C$. This leads to the detuning of the bridge, i.e., non-zero current $I$ is measured by the lock-in. In order for the driving force to oscillate at the same frequency as $u(t)$, the chip is biased by 10~V DC-voltage $U_B$, and the on-resonance driving force is then $F(t) = C_0 U_B u(t)/h$.

The experiment was performed in the helium bath cryostat. The temperature of He~II was controlled by the pumping of saturated vapour together with a resistive heater, whose power was regulated by a PID algorithm. This provided temperature stability with an error $\pm$0.1~mK.

\section{Resonant modes characterization}
\label{sec:modes}

Previous works \cite{varga2020, novotny2024, novotny2024_arxiv} focused on the fundamental Helmholtz acoustic resonant mode. Here, we discuss experimentally and theoretically the higher-lying acoustic modes in the nanofluidic system.

\subsection{Experimentally observed acoustic modes}

\begin{figure}[h!]
    \centering
    \includegraphics[width = \textwidth]{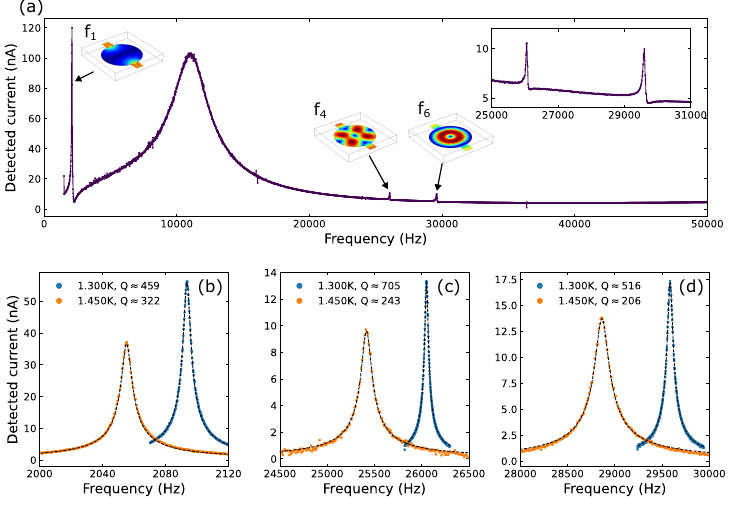}
    \caption{(a) Wide resonant spectrum, the amplitude of the current against the frequency of the driving signal. The resonant peaks are marked by an arrow and corresponding velocity field from the FEM simulation. The inset shows the $f_4$ and $ f_6$ modes in detail. (b-d) Detail of each resonance peak for two different temperatures. The dashed line is the Lorentzian fit.}
    \label{fig:fs}
\end{figure}

Figure \ref{fig:fs}(a) shows a broad spectrum of the total detected acoustic amplitude measured at 1.3~K. The signal was calculated as $I = \sqrt{X^2 + Y^2}$, where $X$ and $Y$ are the two quadratures of the bridge current detected by the lock-in amplifier. The detected current is proportional to the change of the capacitance of the chip $I = U_B dC/dt = U_B (dC/dy) \dot{y}$ \cite{varga2020}, where $\dot{y}$ is the velocity amplitude of the flow. Note, however, that the velocity mode shape of the higher modes has spatial dependence and is not uniform and localised to the inlet channels as for the fundamental Helmholtz mode.

The broad resonance spectrum in Fig. \ref{fig:fs}(a) contains 4 resonant peaks. The peak located around 11~kHz is the RLC resonance of the circuit and is unrelated to the studied physics. The 4th sound acoustic resonances lie around 2100~Hz, 26050~Hz and 29500~Hz and are shown in detail in the Figs. \ref{fig:fs}(b-d) for two different temperatures, both peaks in each frame are measured with similar driving voltage $u(t)$. These individual peaks have been corrected by subtraction of a linear background, which is mainly given by the residual imbalance of the capacitance bridge.

Comparing the measured resonance frequencies with those obtained in the FEM (finite element modelling) simulation, see Fig. \ref{fig:simul1}, we can identify $f_1$ and $f_6$ resonant modes. Nevertheless, in the case of the mode detected at 26050~Hz, $f_4$ or $f_5$ mode can be considered. These modes are nearly degenerate and have similar mode shapes, rotated with respect to each other by 45$^\circ$. The degeneracy of the modes is lifted by flow in the inlet channels. The deformation of the substrate, i.e., the change of the capacitance $\delta C(t)$, is proportional to the pressure fluctuation inside the basin: for positive pressure, the electrodes are repulsed, and for negative pressure, they are attracted. For the total detected current, we have
\begin{equation}
    I = U_B \frac{\dd}{\dd t}(C_0 + \delta C(t)) = U_B \frac{\dd \delta C(t)}{\dd t} \propto U_B \frac{\dd}{\dd t} \left( \int p(\mathbf{r}) \dd S \right),\label{det_curr}
\end{equation}
where the displacement $w(\mathbf{r})$ is proportional to the pressure $p(\mathbf{r})$, and the integration is taken over the basin wall. The integral is from the symmetry trivially zero for the $f_5$-mode since the areas of negative and positive pressure are equal (see Fig.~\ref{fig:simul1}), while for the $f_4$ mode the flow in the channels results in non-zero spatially averaged pressure oscillation. Thus, the detected peak at 26050~Hz corresponds to the $f_4$ resonant mode. It would not be possible to observe this mode if the channels were absent.

From Fig. \ref{fig:fs}(a), the $f_1$ mode has approximately 10x higher amplitude than the higher modes at the same driving voltage. Two effects cause this: First, similarly to the previous case with the peak detection, for $f_1$, the capacitor area deforms only in one direction at a given time during the entire oscillation period, causing either negative or positive pressure change in the whole cavity, i.e., no mutually compensating currents are present. For the $f_4$ and the $f_6$ mode, we have to calculate the integral in the \eqref{det_curr}, and the final detected current will be significantly lower than for the $f_1$ mode. Second, the higher order modes are driven less efficiently by the electrostatic force between the basin electrodes. The effective force $F_{e}$, which induces flow inside the cavity, is calculated as a projection of the electric force acting between basin plates $\mathbf{F} = C_0 U_B u(t)/h \mathbf{\hat{z}}$ on the deformation mode shape $\mathbf{\hat{w}}(\mathbf{r})$ normalized to the unit magnitude:
\begin{equation}
    F_{e} = \frac{1}{S} \int \mathbf{F} \cdot \mathbf{\hat{w}}(\mathbf{r}) \dd S = \frac{C_0 U_B u(t)}{h S} \int \mathbf{\hat{z}} \cdot \mathbf{\hat{w}}(\mathbf{r}) \dd S,\label{Feff}
\end{equation}
where $\mathbf{\hat z}$ is the unit vector in the z-direction. We neglect any side effects and assume the force acts preferentially in the z-direction (perpendicular to the basin surface). Evaluating \eqref{Feff} using the deformation shape obtained from the FEM simulation (see below), we get that at the same voltage, $f_1$ mode is driven by approximately 7x - 8x larger force than $f_4$ and $f_6$ mode.

The quality factors $Q = f_m/\gamma$, where the resonance frequency $f_m$ and the width $\gamma$ are obtained from a fit to $\chi^{-1}(f) = m(f^2 - f_m^2 + if\gamma)$, are shown in the Fig. \ref{fig:fs}(b). For sufficiently low forcing, no turbulence is produced, the resonance is in the linear regime, and the width of the peak is equal to the intrinsic dissipation of the superfluid flow determined by the thermal effects and residual motion of the normal fluid \cite{souris2017}. We observed a drop in $Q$ with increasing temperature, which is mainly caused by the increase of the peak width, i.e., the increase of the intrinsic damping. The resonance frequency changed only by $\approx$ 2\% between both temperatures. The drop in $Q$ is more significant for two higher-lying modes where it is reduced by a factor of $2.5$ between 1.3~K and 1.45~K, which is due to the frequency dependence of the viscous penetration depth $\delta\propto f^{-1/2}$, resulting in increased dissipation due to the residual normal fluid flow at higher frequencies \cite{souris2017}.

\subsection{FEM simulation and analytical description}
\label{sec:modes-calculation}
\begin{figure}[h!]
    \centering
    \includegraphics[width = \textwidth]{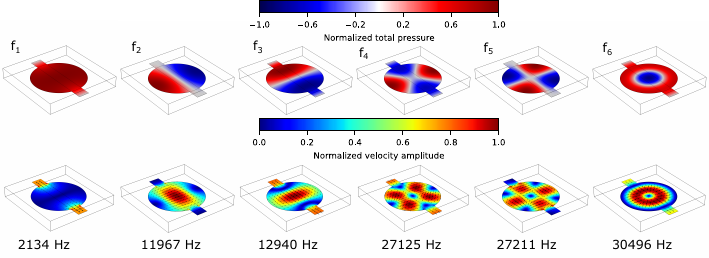}
    \caption{FEM acoustic simulation of the first six resonant modes inside the nanofluidic cavity. The simulation includes acoustic interaction between the fluid and the glass body of the chip, which is marked by the black frame. The first row shows the normalized pressure, and the second row shows the normalized acoustic fluid velocity. Black arrows indicate the direction of the fluid flow.}
    \label{fig:simul1}
\end{figure}

In order to identify the observed resonant modes, we modelled the system using a finite-element (FEM) acoustic simulation using COMSOL simulation software. The substrate (quartz) is treated as a classical linear elastic material and \4He inside the cavity as a linear acoustic medium, including acoustic-structure boundary coupling, which reflects the response of the soft substrate to the moving fluid. The obtained resonance mode shape for the acoustic pressure and velocity are shown in Fig. \ref{fig:simul1}. The resonance frequencies are calculated for the fourth sound velocity 230.9~ms$^{-1}$ \cite{brooks1977}, corresponding to the temperature 1.30~K used in the experiment. The simulation is linear, i.e., the non-linear terms in fluid dynamical equations are omitted, and the velocity field is irrotational. The FEM simulation predicts the frequencies with a relative error between 2\% and 4\%. The discrepancy is likely due to uncertainties in material parameters and imperfect modelling of the superfluid two-fluid behaviour by classical acoustics.

Higher-lying modes are qualitatively different from the fundamental $f_1$ mode: As can be seen in Fig.~\ref{fig:simul1}, the velocity for the $f_1$ mode is localized in the channels, and the presence of the channels is crucial for its existence. For other modes, the flow inside the basin is dominant, and the resonance frequency is approximately given by the acoustic modes of the cylindrical basin.

The analytical derivation of resonance frequencies must also consider the eigenmodes of the mechanically compliant container. To describe the fourth sound acoustic mode in a compliant container, we start from the linearized equations of motion of the superfluid velocity $\mathbf{u}$ inside the device,
\begin{equation}
    \label{eq:euler}
    \frac{\partial \bv u}{\partial t} = -\frac{1}{\rho}\nabla p,
\end{equation}
and of the out-of-plane deformation of the substrate $w$ \cite{reddy2006},
\begin{equation}
    \label{eq:KL} 
    I_0 \frac{\partial^2 w}{\partial t^2} = p - D\nabla^2(\nabla^2 w),
\end{equation}
where $D = (H^3 E)/(12(1 - \nu^2))$ is the flexural rigidity, $H$ is the thickness of the substrate, $E$ the Young modulus, $\nu$ the Poisson ratio, and $I_0 = \rho_\mathrm{sub} H$ with $\rho_\mathrm{sub}$ the density of the quartz substrate. These are supplemented by the equation of continuity that takes into account the displacement of the top and bottom walls and the fact that only the superfluid component moves for the fourth sound,
\begin{equation}
    \label{eq:cont}
    \frac{\partial \rho}{\partial t} = -\rho_s\nabla\cdot\bv u - 2\frac{\rho}{h}\frac{\partial w}{\partial t},
\end{equation}
where we use the directional convention that bowing toward the outside (i.e., increasing volume) is indicated by positive $w$. Differentiating the equation of continuity \eqref{eq:cont} with respect to time and substituting the superfluid Eurler equation \eqref{eq:euler} we arrive at
\begin{equation}
    \frac{\partial^2\rho}{\partial t^2} = \frac{\rho_s}{\rho} \nabla^2p - 2\frac{\rho}{h}\frac{\partial^2 w}{\partial t^2}.
\end{equation}
Finally, using the compressibility relation $\delta\rho = \rho\chi\delta p$, we arrive at
\begin{equation}
    \label{eq:p-eom}
    \frac{\partial^2 p}{\partial t^2} = \frac{\rho_s}{\rho^2\chi}\nabla^2 p - \frac{2}{h\chi}\frac{\partial ^2 w}{\partial t^2},
\end{equation}
which, together with \eqref{eq:KL} form a closed set of equations for the acoustic modes in a compliant container. For our case of a right cylinder of radius $R$ and height $h$, the boundary conditions are $\dd p/\dd r(r = R) = 0$ and $w(r=R) = 0$, $ M_{rr}(r=R) = 0$, which corresponds to the simply supported circular plate. The $M_{rr}$ is the radial part of the plate bending moment \cite{reddy2006}.

For the decoupled motion of the acoustic pressure and the wall, the resonant modes are given by \cite{reddy2006, pierce2019}, in polar coordinates $(r, \theta)$,
\begin{equation}
    \label{eq:p-mode}
    p_{sn}(r, \theta, t) = P J_n\left(\frac{\gamma_{sn}}{R}r\right)\cos(n\theta)\sin(\omega_{sn} t) = P \hat{p}(r, \theta) \sin(\omega_{sn} t)
\end{equation}
and
\begin{align}
\begin{split}
    \label{eq:w-mode}
    w_{sn}(r, \theta, t) &= W\left[J_n\left(\frac{\tilde{\gamma}_{sn}}{R}r\right) - \frac{J_n(\tilde{\gamma}_{sn})}{I_n(\tilde{\gamma}_{sn})}I_n\left(\frac{\tilde{\gamma}_{sn}}{R}r\right)\right]\cos(n\theta)\sin(\omega_{sn} t) \\
    &= W \hat{w}(r, \theta) \sin(\omega_{sn} t)
\end{split}
\end{align}
where $P$ and $W$ are the amplitudes of the pressure and wall deformation oscillation, respectively, $J_n(x)$ is the Bessel function of the first kind and $I_n(x)$ is the modified Bessel function of the first kind. The index $n$ indicates the number of radially extending nodal lines (that is, the azimuthal ``quantum number'') and $s$ ($s+1$ for the pressure mode) is the number of concentric nodal circles (i.e., the radial ``quantum number''). The constants $\gamma_{sn}$, $\tilde{\gamma}_{sn}$ are determined by the boundary conditions, which can be determined separately for the two modes for the decoupled motion. For coupled motion, however, the boundary conditions for pressure and wall displacement differ. As a result, pure pressure eigenmodes will couple to several pure wall eigenmodes and vice versa. Seeking the modification to the acoustic resonance frequency $\omega_{sn}$, we assume to the first order of approximation that $\gamma_{sn}$ is the unperturbed $s$-th zero of the $J'_n(x)$ and that the coupling to other $(sn)$ wall eigenmodes is negligible. Substituting \eqref{eq:p-mode} and \eqref{eq:w-mode} to \eqref{eq:p-eom} and \eqref{eq:KL}, multiplying by the spatial parts $\hat{p}$, $\hat{w}$ and integrating over the area of the circle we obtain 
\begin{subequations}
\label{eq:projected-eq}
\begin{align}
    -\omega_{sn}^2 P |\hat{p}|^2 &= -\frac{\rho_s}{\rho^2\chi}\left(\frac{\gamma_{sn}}{R}\right)^2 P |\hat{p}|^2 + \frac{2}{h\chi}W\omega_{sn}^2 \langle \hat{p},\hat{w} \rangle\\
    -\omega_{sn}^2 I_0 W |\hat{w}|^2 &= P\langle \hat{p}, \hat{w} \rangle - D\left(\frac{\tilde{\gamma}_{ns}}{R}\right)^4W |\hat{w}|^2
\end{align}
\end{subequations}
where $|\hat{p}|^2 = \int_0^{2\pi} \int_0^R \hat{p}^2 r \dd r \dd \theta$, $|\hat{w}|^2 = \int_0^{2\pi} \int_0^R \hat{w}^2 r \dd r \dd \theta$ and $\langle \hat{p}, \hat{w} \rangle = \int_0^{2\pi} \int_0^R \hat{w}\hat{p} r \dd r \dd \theta$. The frequencies are obtained via the condition of existence of non-zero solutions to $P$ and $W$, i.e. $\operatorname{det} M = 0$, where $M$ is the matrix given by the system of linear equations \eqref{eq:projected-eq}. Performing the necessary algebra, we get a biquadratic equation for the resonance frequencies
\begin{equation}
    \label{eq:biquadric}
    I_0 \omega_{sn}^4 + \omega_{sn}^2 \left[ - D\left( \frac{\tilde{\gamma}_{sn}}{R} \right)^4 - I_0 \frac{\rhos}{\rho^2 \chi} \left( \frac{\gamma_{sn}}{R} \right)^2 - \frac{2}{h \chi} C_{sn} \right] + \frac{\rhos}{\rho^2 \chi} D \left( \frac{\gamma_{sn}}{R} \right)^2 \left( \frac{\tilde{\gamma}_{sn}}{R}\right)^4 = 0,
\end{equation}
where $C_{sn} = \langle \hat{p},\hat{w} \rangle^2/(|\hat{p}|^2 |\hat{w}|^2)$ is the normlaized overlap between the pressure and deformation mode. By solving the Eq. \eqref{eq:biquadric}, we obtain two $\omega_{sn+}^2$ and $\omega_{sn-}^2$ as the roots of a quadratic equation (we omit the rather lengthy exact expression). The calculated frequencies are shown in the table \ref{tab:freq}.

\begin{table}[h!]
    \caption{Resonance frequencies calculated based on \eqref{eq:biquadric} for $R = 2.5 \times 10^{-3}$~m, $H = 0.5 \times 10^{-3}$~m, $E = 72$~GPa, $\nu = 0.17$, $\rho_\mathrm{sub} = 2170$~kgm$^{-3}$ and helium properties at 1.300~K from ref.~\cite{donnelly1998}.}
\begin{indented}
    \item[]\begin{tabular}{cccccc}
    \hline
      mode & $\gamma_{sn}$ & $\tilde{\gamma}_{sn}$ & $C_{sn}$ & $f_{-}$ (Hz) & $f_{+}$ (MHz) \\ \hline 
      $f_4$ & 3.054 & 5.06& 0.65 & 27075 & 0.89 \\
      $f_6$ & 3.832 & 5.45 & 0.52 & 39294 & 0.89 \\ \hline
    \end{tabular}
\end{indented}
\label{tab:freq}
\end{table}
The lower lying frequencies $f_{-}$ are the perturbed eigenmodes in the helium due to interaction with the substrate. The frequencies $f_{+}$ lie in the MHz range and correspond to the eigenmodes of the substrate perturbed by the interaction with helium, which are undetectable in our experiment. In Tab.~\ref{tab:freq}, we see that the estimated frequency $f_4$ mode is in excellent agreement with the experiment, but $f_6$ is over-estimated by about 30\%. In the calculation, we assumed only the coupling between modes with the highest value of $C_{sn}$, i.e., the pressure mode is coupled only to the most similar deformation mode. In general, the pressure mode can be coupled to multiple deformation modes, so one should consider deformation in the form of the sum $w = \sum_{sn}W_{sn}\hat{w}_{sn}\sin(\omega t)$ and then obtain a set of equations for amplitudes $W_{sn}$ by projection on each separate eigenmode $\hat{w}_{sn}$. This would lead to the set of coupled equations similar to \eqref{eq:projected-eq}. The single-mode approximation is sufficient for the $f_4$ mode, where the overlap integrals with $(s=0, n=2)$ wall mode $C_{02}\approx 0.64$, $C_{12}\approx 0.1$ and $C_{22}\approx 0.05$, i.e., there is a dominant coupling to a single mode. For the $f_6$ mode, however, $C_{00}\approx0.27$, $C_{10}\approx0.52$ and $C_{20}\approx0.08$, and these modes in turn couple to multiple pressure modes, which is likely the reason for over-estimation of the $f_6$ mode resonance frequency.
 
\section{Turbulence-mediated dissipative coupling between modes}
\label{sec:coupling}

\begin{figure}[h!]
    \centering
    \includegraphics[width= \textwidth]{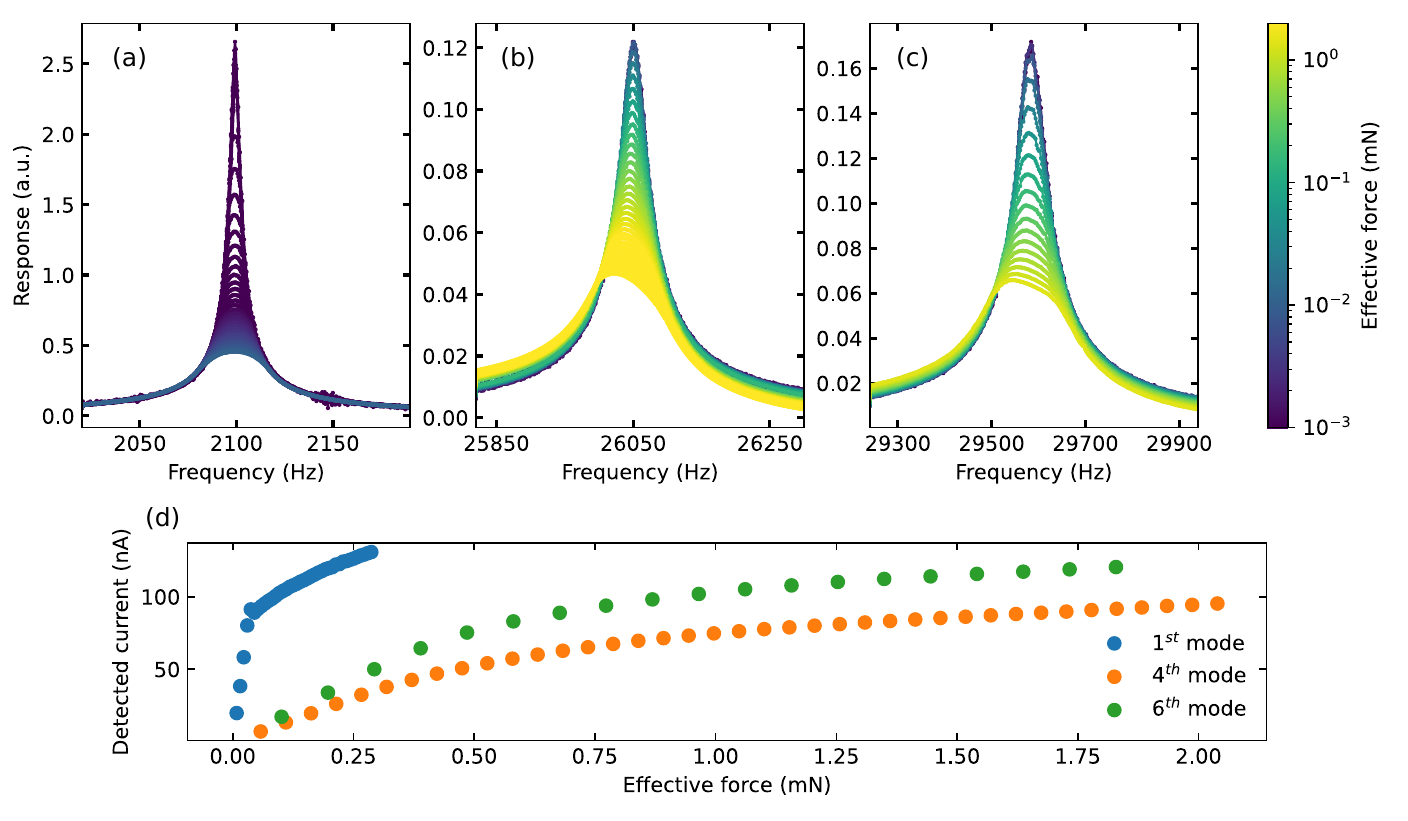}
    \caption{(a-c) Detected current divided by the effective force against the frequency for a range of applied effective force at 1.30~K in the logarithmical colour scale. Peaks are corrected with respect to the linear background, which is extrapolated from a set of linear peaks at low forcing. Note that the colour scale is the same for the three cases; nonlinear dissipation is significantly more prominent for the fundamental mode. (d) The amplitude of the detected current against the effective force.}
    \label{fig:resp_vs_force}
\end{figure}

As the driving force in the cavity increases beyond a critical threshold, the dissipation becomes non-linear due to the generation of quasi-2D turbulence consisting of quantized vortices \cite{barenghi2023,varga2020,sachkou2019}. This transition introduces an additional energy dissipation mechanism due to mutual friction, which couples the relative motion of the superfluid and normal fluid components and quantized vortices \cite{Varga2019}. Consequently, the amplitude of the measured resonance peaks ceases to scale linearly with the applied forcing, as shown in Fig. \ref{fig:resp_vs_force}(a-c), where we plot a detected response divided by the effective force \eqref{Feff}. The shape of the resonance peaks deviates from the typical Lorentzian profile. In Fig. \ref{fig:resp_vs_force}(d), we show the measured amplitude of the resonance peak corrected with respect to the linear background against the effective force. The transition to turbulence of the $f_1$ mode is sharp and abrupt, consistent with hysteretic turbulent behaviour observed previously \cite{varga2020,novotny2024_arxiv}. The transition in the case of the $f_4$ and $f_6$ mode is much smoother and occurs at almost two orders higher effective force. This difference is most likely due to a relatively small fraction of the kinetic energy of the flow that is concentrated near the sharp corners between the basin and the inlet channels (see Sec.~\ref{sec:local}), making the modes less sensitive to the appearance of turbulence.

\begin{figure}[h!]
    \centering
    \includegraphics[width = \textwidth]{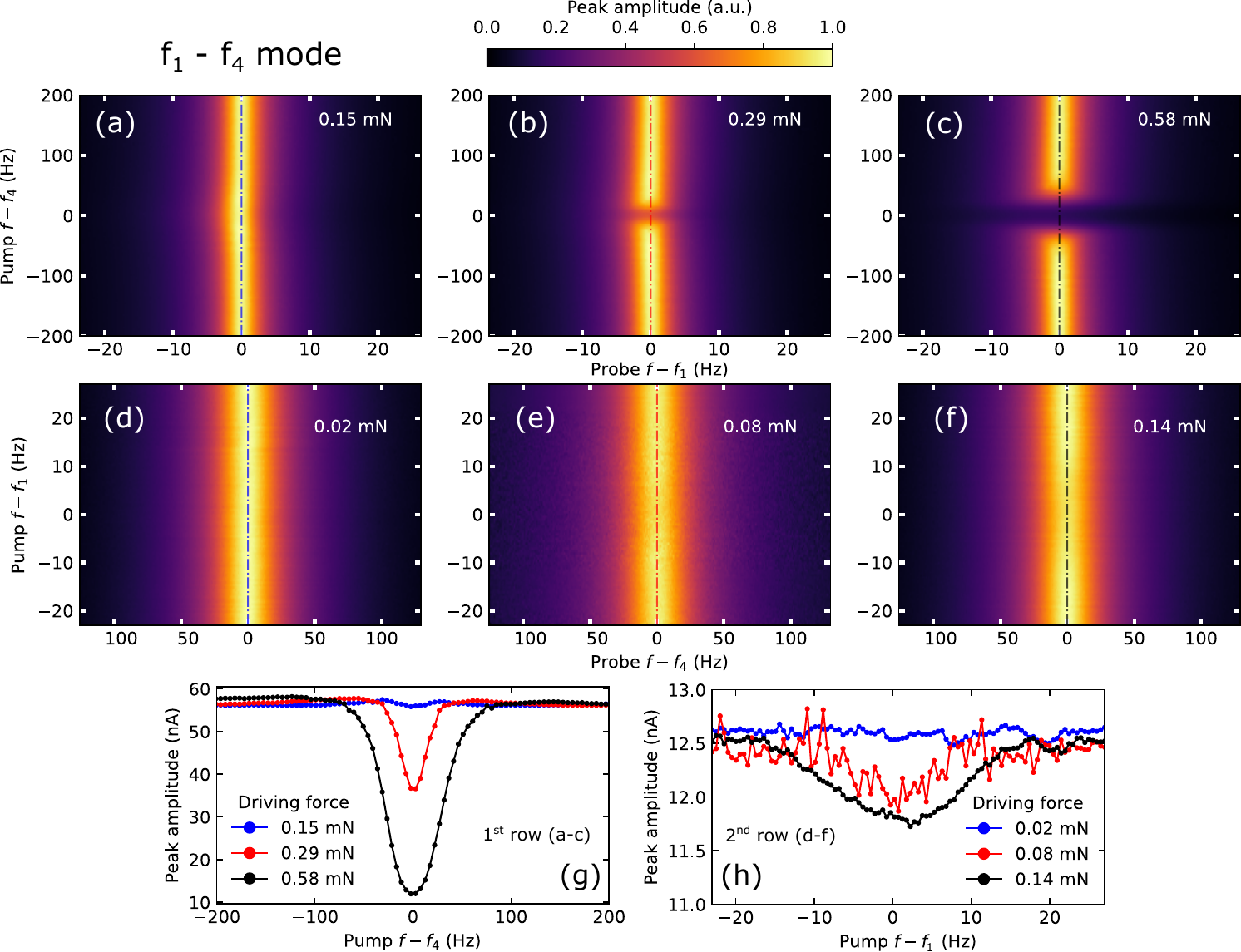}
    \caption{Heatmaps of the interaction between the $f_1$ and $f_4$ mode. (a-c) $f_1$ mode is used as the probe. (d-f) $f_4$ mode is used as the probe. The effective force of the drive mode is displayed in the top right corner of each image. The dashed-dot line in each frame symbolizes the trace, along which the cut (g-h) through the heatmap is made.}
    \label{fig:500A_2D_f1}
\end{figure}

\begin{figure}[h!]
    \centering
    \includegraphics[width = \textwidth]{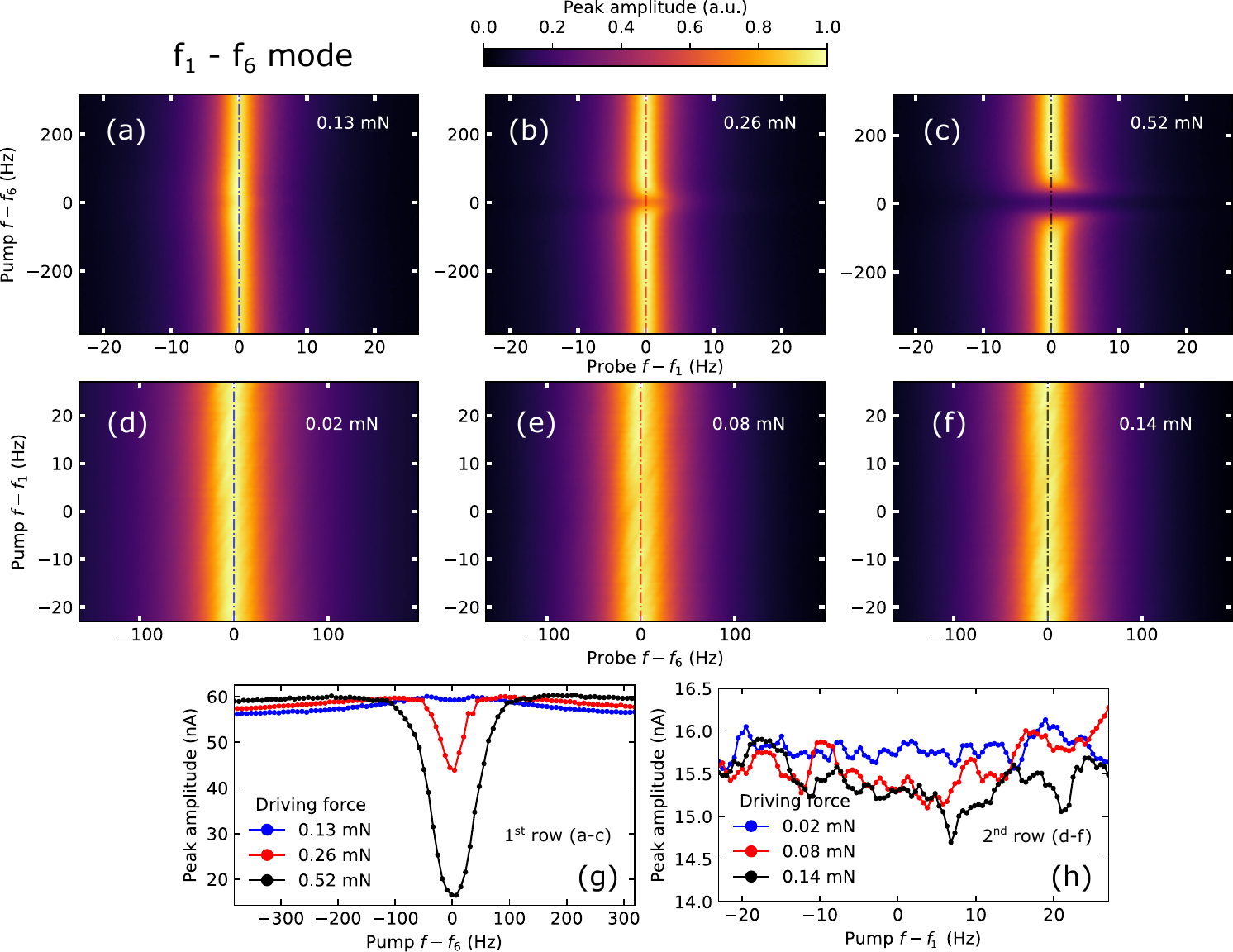}
    \caption{Heatmaps of the interaction between the $f_1$ and $f_6$ mode. (a-c) $f_1$ mode is used as the probe. (d-f) $f_6$ mode is used as the probe. The effective force of the drive mode is displayed in the top right corner of each image. The dashed-dot line in each frame symbolizes the trace, along which the cut (g-h) through the heatmap is made.}
    \label{fig:500A_2D_f2}
\end{figure}

We studied a mutual interaction between the fundamental and two higher modes when one of the modes was driven in the linear regime (the \emph{probe mode}) and the second in the turbulent regime (the turbulent \emph{pump mode}), similar to the experiment \cite{midlik2021}, where two modes of second sound in a bulk three-dimensional cavity were used. In the Figs. \ref{fig:500A_2D_f1} and \ref{fig:500A_2D_f2}, we show heatmaps, where the colour symbolizes the strength of the probe signal. The X-axis shows the frequency of the probe signal, which was continuously swept, and the Y-axis shows the frequency of the pump, which was changed in steps between sweeps across the probe peak. The amplitude of the fundamental mode (as probe) was attenuated when we approached the resonance frequency, $f_4$ or $f_6$ (as pumps), of a higher mode at sufficiently high forcing. This can be seen in the Figs. \ref{fig:500A_2D_f1}(g) and \ref{fig:500A_2D_f2}(g), which show a vertical cut through the heatmap at the resonance frequency of the fundamental $f_1$ mode. For the highest driving force, the higher mode resonant peak is replicated in the dissipation of the fundamental mode, which is reduced by nearly 80\% in the case of $f_1$-$f_4$ mode interaction. The induced attenuation of the probe mode is caused by the mutual friction force acting between the fourth sound resonance and quantized vortices \cite{novotny2024} generated by the flow due to the pump mode, which is in the nonlinear regime for the force amplitudes that show the induced attenuation (Fig. \ref{fig:resp_vs_force}(b,d)). The critical velocity for the appearance of nonlinear damping and the dissipative interaction is likely given by the depinning velocity \cite{novotny2024_arxiv, schwarz1985}, although a direct comparison, as for the fundamental Helmholtz mode in \cite{novotny2024_arxiv} is not straightforward due to the non-uniformity of the flow.

Surprisingly, analogous strong attenuation was not noticed in the reverse case when one of the higher modes was used as a probe, even though the $f_1$-pump mode was driven with the maximum possible force given by experimental limitations. We observed only minor attenuation of the higher mode, at most by approximately $7\%$ for the probe by the $f_4$ mode, see Fig. \ref{fig:500A_2D_f1}(h), and no discernible attenuation of the $f_6$ mode, see Fig.~\ref{fig:500A_2D_f2}(h).

\subsection{Turbulence detection by the $f_1$ mode}
\begin{figure}[h!]
    \centering
    \includegraphics[width = \textwidth]{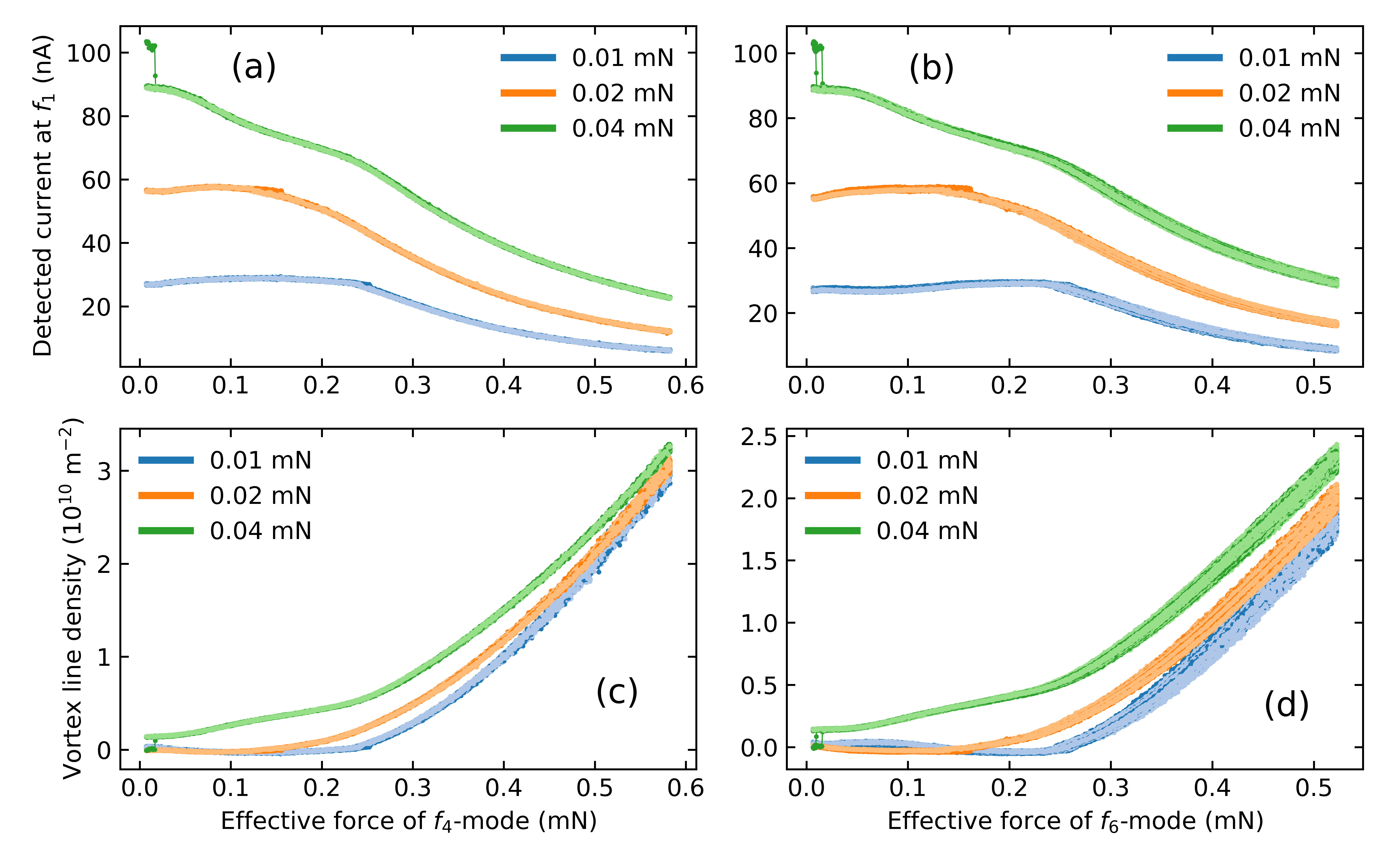}
    \caption{(a-b) Dependence of the $f_1$ mode amplitude on the higher mode effective driving force. (c-d) Vortex line density detected by the $f_1$ mode against the higher mode effective driving force.}
    \label{fig:as_Ls}
\end{figure}

The dependence of the $f_1$ mode amplitude on the pump forcing of a higher mode on resonance is displayed in Figs. \ref{fig:as_Ls}(a,b), the dark part of the curve symbolizes the gradual increase of the driving force, while the light part of the curve shows the decrease. In Figs. \ref{fig:as_Ls}(c,d), we show the average amount of quantized vortices per unit area (vortex line density $L$) calculated from the attenuation of the fundamental mode using the formula~\cite{novotny2024}
\begin{equation}
    L = \frac{2 \pi \gamma_0}{\alpha \kappa}\left( \frac{a_0}{a} - 1\right),
\end{equation}
where $a_0$ and $a$ is the unattenuated and attenuated amplitude, $\alpha$ the mutual friction coefficient \cite{donnelly1998} and $\gamma_0$ the width of the unattenuated peak. This formula assumes that vortices are stretched between the top and bottom walls of the nanofluidic cavity and that the superfluid flow is perpendicular to them \cite{novotny2024}. 

We measured the dependence of $L$ on the pump mode forcing for three different excitations of the $f_1$ probe mode. Flows corresponding to the 0.01~mN and 0.02~mN probe excitation were in the linear regime, while the flow at 0.04~mN was close to the turbulent transition. This also manifests in the presence of bistable behaviour \cite{varga2020,novotny2024_arxiv}, i.e., the discontinuous jumps in the amplitude for a few dark green curves. The probe flow existed in a meta-stable laminar state, and a perturbation in the form of the higher mode flow caused an abrupt collapse to a turbulent state. Similar behaviour, but less prominent, was seen in the case of curves measured at 0.02~mN (orange curves) around 0.15~mN of the effective pumping force. In general, the probe signal was almost constant or slightly amplified (see \ref{app:att}) for lower pump forcing. At higher drives, we observed a drop due to the onset of turbulence, which is also apparent from the increase of the vortex line density, which is consistent with the frequency sweeps in Figs.~\ref{fig:500A_2D_f1}, \ref{fig:500A_2D_f2}.

Furthermore, the critical effective force of a higher mode, at which the flow transitions to turbulence, decreases with increasing probe $f_1$ mode driving. That is, turbulence does not develop independently for the two modes. The sum of the probe and pump force amplitudes at the critical point is approximately constant, suggesting a critical velocity, localized likely near the sharp corners connecting the basin and channels, that is independent of the mode shape.

\subsection{Sensing of localized turbulence}
\label{sec:local}
The vortex line density, which is detected by a resonant mode, is an average over the whole domain weighted by the velocity profile of the given mode. The average vortex line density $\langle L \rangle$ is given by\cite{varga2015}
\begin{equation}
    \label{eq:mean_L}
    \langle L \rangle = \frac{\int L (\mathbf{r}) |\mathbf{u}(\mathbf{r})|^2 \dd V}{\int |\mathbf{u}(\mathbf{r})|^2 \dd V},
\end{equation}
where $L(\mathbf{r})$ is the spatial distribution of quantized vortices in the cavity. The highest vortex line density will be concentrated near regions of strong shear, i.e., close to the sharp corners at the connection between the channel and the basin. This holds especially for the $f_1$ mode, whose velocity field is suppressed everywhere except the channels. The velocity field of the higher modes inside the basin is irrotational, with the strongest shear concentrated near the sharp corners (see Fig.~\ref{fig:simul1}). This suggests that the initial development of turbulence is concentrated near these sharp edges for all modes. This provides an explanation for the asymmetry of the turbulent coupling: 

The velocity of the fundamental mode is concentrated in the inlet channels (see Fig.~\ref{fig:simul1}) and will thus have a strong overlap integral between $L(\bv r)$ and $|\bv u(\bv r)|^2$ in \eqref{eq:mean_L}, resulting in the strongly damped fundamental mode due pumping of the higher modes.

In the reverse experiment, i.e., pump on the fundamental $f_1$ and probing on the higher $f_4$ and $f_6$, we observed only minor attenuation. The fundamental mode will generate turbulence, which is again localised in the channels. However, the overlap integral \eqref{eq:mean_L} will be significantly less. The velocities $\mathbf{u_{4,6,b}}$ in the basin can be obained by plugging the Eq. \eqref{eq:p-mode} into \eqref{eq:euler}, the exact formulas are shown in the \ref{app:modes}. The velocity in the channel is taken as spatially uniform and perpendicular to the channel cross-section
\begin{equation}
    \label{v_channel}
    \mathbf{u_{4,6,ch}} = \dfrac{1}{\rho \omega}\dfrac{p_{4,6}(R,0,t)}{l} \hat{\mathbf{x}},
\end{equation}
where $p_{4,6}(R,0,t)$ is the pressure at the edge of the basin, and $l$ is the length of the channel. To a first approximation, we take the pressure to be linearly decreasing towards the end of the channel, where it is taken to be zero. The energy is calculated as $E = 1/2 \int \rhos |\mathbf{u}(\mathbf{r})|^2 \dd V$ and $\rhos$ is uniform everywhere, thus we get the ratio between the energy in the basin $E_b$ and the channel $E_{ch}$
\begin{equation}
    \frac{E_\mathrm{b}}{E_\mathrm{ch}} = \dfrac{\int |\mathbf{u_b}|^2 \dd V}{\int |\mathbf{u_{ch}}|^2 \dd V}.\label{E_rat}
\end{equation}
Calculating the integrals over the basin velocity shape modes $\mathbf{u_{4,6,b}}$ and using \eqref{v_channel}, we obtain the ratios for both higher modes $E_\mathrm{b}/E_\mathrm{ch} = 4.39$ for the $f_4$ mode and 6.41 for $f_6$. The channel is included twice in the calculation to reflect the resonator geometry.

This indicates that the overall sensitivity of the superfluid flow on quantized vortices is more than 4x larger in the basin. The sensitivity is proportional to the energy due to quadratic dependence on velocity. Further, the ratio is greater for the $f_6$ mode, which is in agreement with the observation of no measurable attenuation by the $f_6$ mode in contrast to the weak $f_4$ mode detection, see Figs. \ref{fig:500A_2D_f1}(h),\ref{fig:500A_2D_f2}(h). It should also be emphasized that the velocity estimate \eqref{v_channel} is not fully correct and leads to the higher values of $E_{\mathrm{ch}}$, i.e. see velocity fields from the FEM simulation Fig.\ref{fig:simul1}, where the velocity in channels is not spatially uniform.

\section{Conclusion}
\label{sec:conclusions}
The main focus of this work is the investigation of localized quantized vortex clusters in quasi-2D flow inside a nanofluidic superfluid fourth sound acoustic resonator used previously to study two-dimensional turbulence \cite{varga2020,novotny2024_arxiv}. In the previous studies, no information was available about the spatial distribution of the vortex line density, but it was assumed that it is localised in the areas of strong superfluid velocity. In this work, we provided experimental support for this assumption by exploring the higher-order resonant modes of fourth sound, which have been largely unexamined in prior studies.

Using finite element method (FEM) simulations and analytical calculation, we identified the spatial distributions of pressure and velocity fields associated with these modes and studied the inter-mode coupling due to turbulence. Specifically, we use the fundamental $f_1$ mode as a turbulence probe while a higher mode serves as the turbulence pump. We observe attenuation of the fundamental mode amplitude as the turbulence pump increases. This attenuation corresponds to the density of quantized vortices generated by sufficiently strong driving of the pump mode. Conversely, when we use a higher mode as the probe and the fundamental as the pump mode, a similar level of attenuation does not occur. We attribute this asymmetry to the differing spatial sensitivities of the resonant modes to quantized vortices: the fundamental $f_1$ mode is primarily sensitive in the channel region, while the higher modes have greater sensitivity in the basin, where vortex density generated by the $f_1$ mode is minimal.

The present work confirms the assumption that the two-dimensional turbulence in acoustically induced flows develops locally in areas of strong shear. Furthermore, the pump-probe measurement technique introduced here will allow future measurements of the dynamics of two-dimensional turbulence, i.e., its development, spatial distribution or decay, which was thus far studied in quantum fluids with only a limited number of quantized vortices \cite{sachkou2019,Gauthier2019}.

\section{Acknowledgments}
We are grateful to \v{S}. Midlik for assistance with the initial stages of device fabrication and L. Skrbek for fruitful discussions. The work was supported by Charles University under PRIMUS/23/SCI/017. CzechNanoLab projects LM2023051 and LNSM-LNSpin funded by MEYS CR are gratefully acknowledged for the financial support of the sample fabrication at CEITEC Nano Research Infrastructure and LNSM at FZU AV\v{C}R. F. N., in addition, acknowledges the financial support from Charles University under GAUK 129724.


\appendix

\section{Mutual signal amplification through detection nonlinearity.}\label{app:att}

\begin{figure}[h!]
    \centering
    \includegraphics[width=0.8\textwidth]{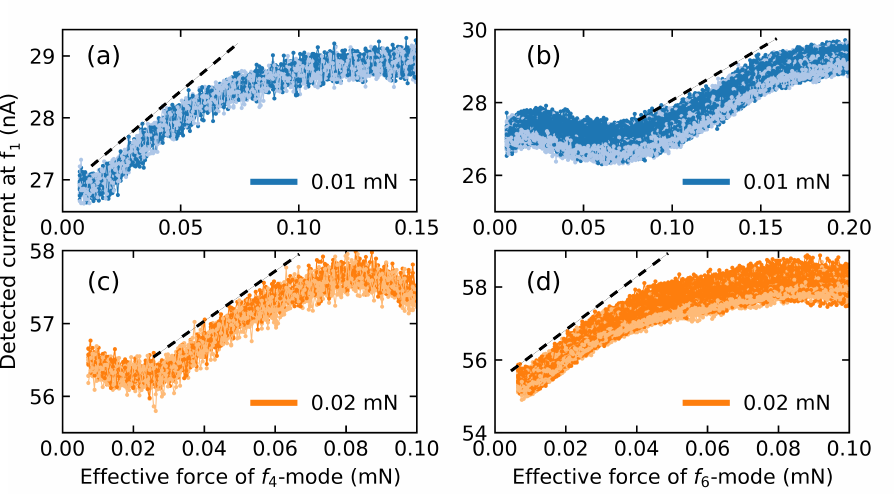}
    \caption{Details of the amplification of the $f_1$ signal at two lowest driving forces due to the interaction with a higher mode, details of the Fig. \ref{fig:as_Ls}(a-b). The dashed line marks the linear part of the amplification.}
    \label{fig:amlifi}
\end{figure}

Besides the attenuation, we observed a slight increase in the amplitude of the linear fundamental mode for low driving effective forces of a higher mode, see Fig.\ref{fig:amlifi}(a-d). In the main part, the increase of the amplitude is linear with the force (in the figures indicated by a dashed line). After that, it saturates due to the onset of attenuation on quantized vortices, which has the opposite effect to the amplification. The same effect is evident also in the cuts of the heatmaps, see two lowest driving forces (blue and red curves) in Figs. \ref{fig:500A_2D_f1}(g),\ref{fig:500A_2D_f2}(g). The amplification is likely caused by the non-linearity of the chip capacitance $C$, similar to the experiment with a nanobeam \cite{cho2018}. 

The detected current is given by the change of the chip capacitance $I = U_B dC/dt$, neglecting any side-effects, we can consider the chip as a plate capacitor $C = \varepsilon_0 S/(h - 2x)$. We expand this into the Taylor series up to the 3rd order with respect to the substrate deformation $x \ll h$
\begin{equation}
    C = \frac{\varepsilon_0 S}{(h - 2x)} \approx \frac{\varepsilon_0 S}{h} + \frac{2 \varepsilon_0 S}{h^2}x + \frac{4\varepsilon_0S}{h^3}x^2 + \frac{8 \varepsilon_0 S}{h^4}x^3. \label{C_expan}
\end{equation}
Further, we can assume that the deformation $x$ can be split into two contributions from two different resonant modes $\omega_1$ and $\omega_2$
\begin{equation}
    x = x_1 + x_2 = x_{01} \cos(\omega_1 t) + x_{02} \cos(\omega_2 t), \label{com_def}
\end{equation}
where $x_{10}$ and $x_{20}$ is the spatial averaged deformation of the given mode. Plugging \eqref{com_def} to the \eqref{C_expan} with the assumption that the lock-in is referenced to $\omega_1$ we get for the detected current
\begin{equation}
    I = U_B \frac{d}{dt} \left[ \frac{2\varepsilon_0 S}{h^2} x_{10} \cos(\omega_1 t) + \frac{8\varepsilon_0 S}{h^4}\left( \frac{3}{4}x_{10}^3 \cos(\omega_1 t) + \frac{3x_{10} x_{20}^2}{2}\cos(\omega_1 t) \right) \right]. \label{eq_mixing}
\end{equation}
The last term on the right-hand side of \eqref{eq_mixing} should be responsible for amplifying the fundamental mode by the $\omega_2$ mode. We are aware of the fact that the relation between the deformation $x_{02}$ and the higher mode driving force is not clear and cannot directly explain the experimentally observed linear increase.

\section{Velocity shape modes}\label{app:modes}
The $f_4$ mode velocity with $\gamma_{02} = 3.054$ is described by equations
\begin{align}
v_{r4}(r,\theta,t) &= \dfrac{P}{\rho \omega_{02}}\dfrac{\gamma_{02}}{R} J_2'\left(\dfrac{\gamma_{02}}{R}r\right)\cos(2\theta)\cos(\omega_{02}t)\\
v_{\theta 4}(r,\theta,t) &= -\dfrac{P}{\rho \omega_{02}}\dfrac{2}{r} J_2 \left(\dfrac{\gamma_{02}}{R}r\right) \sin(2 \theta)\cos(\omega_{02}t),    
\end{align}
and the $f_6$ mode with $\gamma_{10} = 3.832$ by
\begin{align}
    v_{r6}(r,\theta,t) &= \dfrac{P}{\rho \omega_{10}} \dfrac{\gamma_{10}}{R} J_0'\left( \dfrac{\gamma_{10}}{R}r \right)\cos(\omega_{10}t)\\
    v_{\theta 6}(r, \theta,t) &= 0
\end{align}

\begin{figure}[h!]
    \centering
    \includegraphics[width=0.5\linewidth]{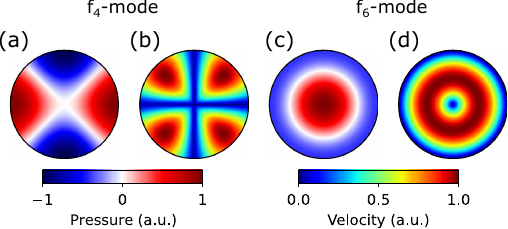}
    \caption{Spatial distribution of velocity and pressure mode shape of the $f_4$ and $f_6$ mode given by analytical expressions.}
    \label{fig:bessel}
\end{figure}

\end{document}